\documentclass[aps,pra,reprint,superscriptaddress,showpacs]{revtex4-1}

\usepackage{graphicx}
\usepackage{amsmath}
\usepackage{float}

\begin{document}


\title{The magnetic and electric transverse spin density of spatially confined light}



\author{Martin Neugebauer}
\affiliation{Max Planck Institute for the Science of Light, Staudtstr. 2, D-91058 Erlangen, Germany}
\affiliation{Institute of Optics, Information and Photonics, University Erlangen-Nuremberg, Staudtstr. 7/B2, D-91058 Erlangen, Germany}
\author{J\"org Eismann}
\affiliation{Max Planck Institute for the Science of Light, Staudtstr. 2, D-91058 Erlangen, Germany}
\affiliation{Institute of Optics, Information and Photonics, University Erlangen-Nuremberg, Staudtstr. 7/B2, D-91058 Erlangen, Germany}
\author{Thomas Bauer}
\affiliation{Department of Quantum Nanoscience, Kavli Institute of Nanoscience, Delft University of Technology, Lorentzweg 1, 2628 CJ Delft, Netherlands}
\author{Peter Banzer}
\email[]{peter.banzer@mpl.mpg.de}
\homepage[]{http://www.mpl.mpg.de/}
\affiliation{Max Planck Institute for the Science of Light, Staudtstr. 2, D-91058 Erlangen, Germany}
\affiliation{Institute of Optics, Information and Photonics, University Erlangen-Nuremberg, Staudtstr. 7/B2, D-91058 Erlangen, Germany}


\date{\today}

\begin{abstract}

When a beam of light is laterally confined, its field distribution can exhibit points where the local magnetic and electric field vectors spin in a plane containing the propagation direction of the electromagnetic wave. The phenomenon indicates the presence of a non-zero transverse spin density. Here, we experimentally investigate this transverse spin density of both magnetic and electric fields, occurring in highly-confined structured fields of light. Our scheme relies on the utilization of a high-refractive-index nano-particle as local field probe, exhibiting magnetic and electric dipole resonances in the visible spectral range. Because of the directional emission of dipole moments which spin around an axis parallel to a nearby dielectric interface, such a probe particle is capable of locally sensing the magnetic and electric transverse spin density of a tightly focused beam impinging under normal incidence with respect to said interface. We exploit the achieved experimental results to emphasize the difference between magnetic and electric transverse spin densities.
\end{abstract}

\pacs{03.50.De, 42.25.Ja, 42.50.Tx}

\maketitle

\section{Introduction}\enlargethispage{4\baselineskip}
The transverse spin density (TSD) of light describes field vectors, which spin transversely with respect to the local propagation direction of the electromagnetic wave~\cite{Aiello2015,Bliokh2015}. In nature, such polarization states occur when electromagnetic waves experience strong lateral confinement, since the appearance of transverse spin is intimately linked to the presence of longitudinal field components~\cite{Aiello2015,Bliokh2015}. Typical optical systems exhibiting a TSD are waveguide modes~\cite{Petersen2014,Sollner2015,Young2015,Lodahl2017}, surface plasmon polaritons~\cite{Bliokh2012,Kim2012,Canaguier-Durand2014}, near fields of nano-strucures~\cite{Saha2016}, whispering gallery modes~\cite{Junge2013} and tightly focused beams~\cite{Banzer2013,Neugebauer2015,Chen2017}.

In recent years, a wide variety of potential applications led to a continuously increasing interest in the TSD (see for instance refs.~\cite{Aiello2015,Bliokh2015,Lodahl2017} and references therein), particularly due to a related directional emission and coupling effect~\cite{Rodriguez-Fortuno2013}. The phenomenon, which is often referred to as spin-momentum locking~\cite{VanMechelen2016,Bliokh2015q}, can be used to implement spin dependent signal routing~\cite{Rodriguez-Fortuno2013,Neugebauer2014,LeFeber2015}, and single atom optical devices such as isolators and circulators~\cite{Sayrin2015a,Scheucher2016}. Thus, the TSD constitutes the foundation for novel quantum information processing concepts at the nano-scale~\cite{Lodahl2017,Sollner2015}. This interest in the TSD also led to the development of highly sensitive measurement techniques, capable of sensing the TSD in propagating and evanescent waves~\cite{OConnor2014,Neugebauer2015}. 

Although the experimental techniques introduced in refs.~\cite{OConnor2014,Neugebauer2015} are mainly concerned with the TSD of the electric field, from a theoretical point of view, both magnetic and electric components contribute equally to the total spin density $\mathbf{s}$~\cite{Aiello2015,Bliokh2015}:
\begin{align}\label{eq:s}
\mathbf{s}&=\operatorname{Im}\left[\mu_{0}\mathbf{H}^{*}\times\mathbf{H}+\epsilon_{0}\mathbf{E}^{*}\times\mathbf{E}\right]/4\omega 
\equiv\mathbf{s}_{H}+\mathbf{s}_{E}\text{,}
\end{align}
where $\omega$ refers to the angular frequency of the time-harmonic wave, $\mathbf{H}$ and $\mathbf{E}$ denote the magnetic and the electric fields, and $\mu_{0}$ and $\epsilon_{0}$ represent the permeability and the permittivity in vacuum. This equally weighted split into $\mathbf{s}_{E}$ and $\mathbf{s}_{H}$ is often referred to as dual symmetry~\cite{Bliokh2014} or electromagnetic democracy~\cite{Berry2009}. While in the highly symmetric case of a single circularly polarized plane wave, the spin density is purely longitudinal and the magnetic and electric components are equal, $s_{H}^{z}= s_{E}^{z}$~\cite{Bekshaev2007,Berry2009,Aiello2015}, in more general fields of light, this equivalence of $\mathbf{s}_{H}$ and $\mathbf{s}_{E}$ does not hold.

Here, we explore both theoretically and experimentally the fundamental difference between the TSD of the magnetic and the electric field. At first, we theoretically elaborate on the distribution and composition of the TSD in the simplified exemplary scenario of a linearly polarized Gaussian beam. Then, we experimentally investigate the TSD of the magnetic and the electric field in tightly focused beams of light. While for the electric TSD, a suitable measurement technique has been presented in ref.~\cite{Neugebauer2015}, the magnetic component of the TSD has, to the best of our knowledge, not been experimentally studied so far. On these grounds, we will detail for the first time a versatile experimental approach for reconstructing the TSD of the magnetic field at the nano-scale, which at the same time allows to access the TSD of the electric field. Finally, we apply the technique to three different tightly focused polarization tailored beams of light and compare the reconstructed components of the respective TSD.

\section{Magnetic and electric transverse spin}
We begin the discussion by exemplarily considering a paraxial linearly $x$-polarized monochromatic Gaussian beam of light, whose electric field distribution can be approximated by~\cite{Mandel1995}
\begin{align}\label{eq:Gauss}
\textbf{E}\left(x,y,z\right)\approx E_{0}\frac{z_{0}\textbf{e}_{x} }{z_{0}+\imath z}  \exp{\left[\imath k z-\frac{kr^2}{2z_{0}+\imath 2z}\right]} \text{,}
\end{align}
where $z_{0}$ and $E_{0}$ represent the Rayleigh range and amplitude of the beam, with \mbox{$r=\left(x^{2}+y^{2}\right)^{1/2}$} as radial coordinate. Evidently, such a field distribution does not fulfill the transverse constraint of Maxwell's equations --- Gauss's law in vacuum~\cite{Jackson1999} --- $\nabla\cdot\textbf{E}=0$. However, it is possible to revise Eq.~\eqref{eq:Gauss} accordingly by introducing a longitudinal field component~\cite{Novotny2006}. In the focal plane ($z=0$), a suitably adapted field distribution can be written as~\cite{Aiello2015}
\begin{align}\label{eq:Gauss_ExEz}
\mathbf{E}\left(x,y\right)\approx E_{0}\left(\textbf{e}_{x} +\frac{\imath x \textbf{e}_{z}}{z_{0}}\right)\exp{\left[ -k\frac{r^2}{2z_{0}}\right]} \text{.}
\end{align}
Following this line of arguments, we can derive a similar expression for the focal distribution of the magnetic field of the described Gaussian beam. By starting with a $y$-polarized magnetic field --- perpendicular to the $x$-polarized electric field --- and by applying Gauss's law of the magnetic field~\cite{Jackson1999}, $\nabla\cdot\textbf{H}=0$, we result in
\begin{align}\label{eq:Gauss_HyHz}
\mathbf{H}\left(x,y\right)\approx H_{0}\left(\textbf{e}_{y} +\frac{\imath y \textbf{e}_{z}}{z_{0}}\right)\exp{\left[ -k\frac{r^2}{2z_{0}}\right]} \text{,}
\end{align}
with $H_{0}=\sqrt{\epsilon_{0}/\mu_{0}}E_{0}$. It is important to note that Eqs.~\eqref{eq:Gauss_ExEz}~and~\eqref{eq:Gauss_HyHz} represent approximations and are only valid for paraxial or weakly focused Gaussian beams of light. However, the equations contain several important features, which illustrate the central message of this letter; a nonzero phase difference between longitudinal and transverse field components, and differing spatial distributions of the magnetic and the electric TSD.

\begin{figure} 
  \includegraphics[width=0.45\textwidth]{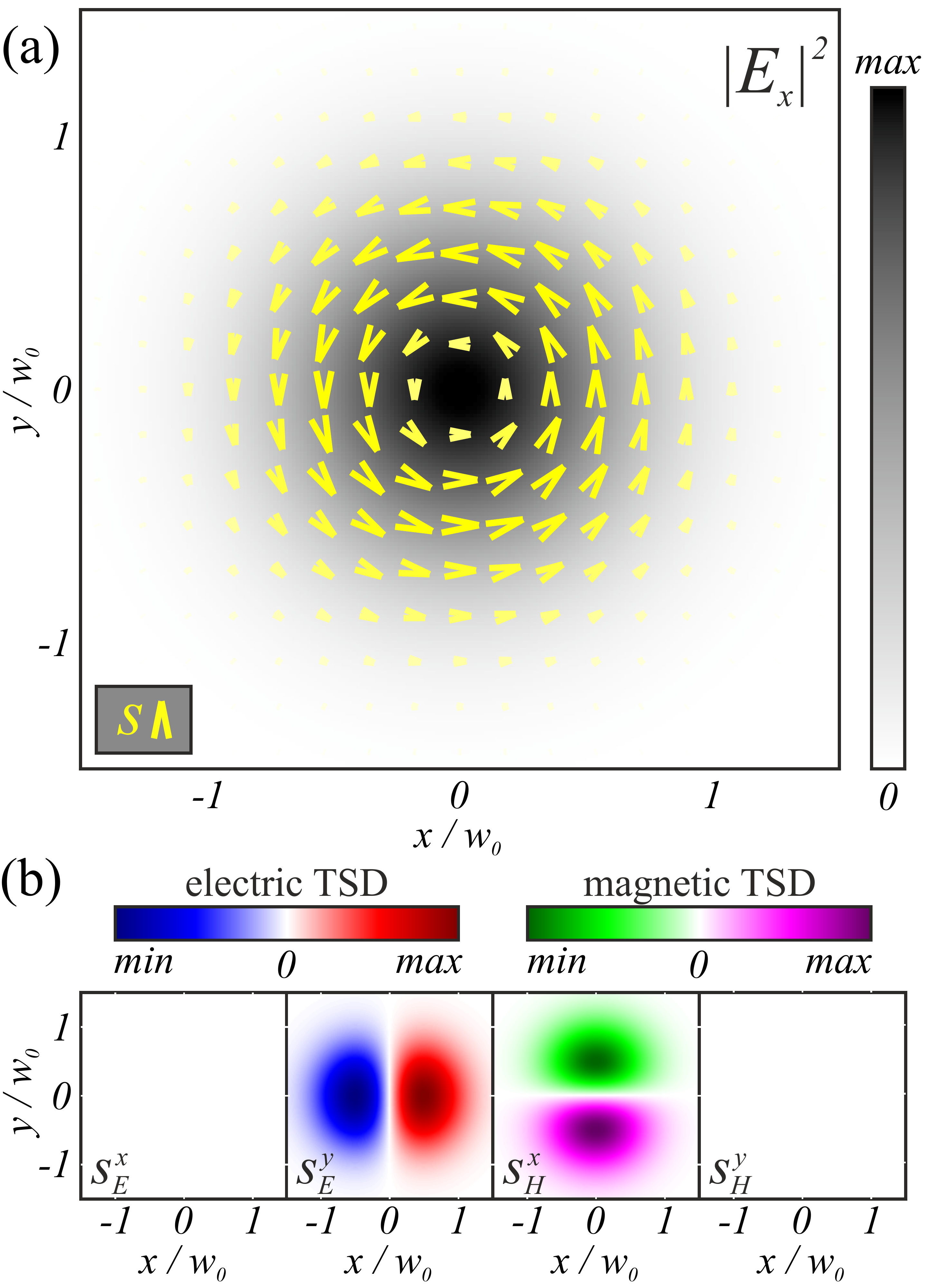}
  \caption{Transverse spin density (TSD) of a linearly polarized Gaussian beam. (a) depicts the energy density $w$ and the spin density $\mathbf{s}$ (yellow arrowheads). (b) illustrates the transverse $x$- and $y$-components of $\mathbf{s}_{E}$ and $\mathbf{s}_{H}$.}
  \label{fig:Gauss}
\end{figure}

At first, we elaborate on the relative phases of the individual field components. As indicated by the imaginary units in Eqs.~\eqref{eq:Gauss_ExEz}~and~\eqref{eq:Gauss_HyHz}, the longitudinal field components are $\pm\pi/2$ out of phase with respect to the transverse field components, resulting in transversely spinning magnetic and electric field vectors wherever the corresponding field components $H_{y}$ and $H_{z}$ or $E_{x}$ and $E_{z}$ overlap~\cite{Aiello2015}. For further investigation of the resulting transverse spin, we calculate the focal spin density distribution by inserting Eqs.~\eqref{eq:Gauss_ExEz}~and~\eqref{eq:Gauss_HyHz} in Eq.~\eqref{eq:s} and yield
\begin{align}\label{eq:Gauss_s}
\mathbf{s}&\approx \left(\epsilon_{0}E_{0}^{2}x\textbf{e}_{y} - \mu_{0}H_{0}^{2}y\textbf{e}_{x}\right)\frac{\exp{\left(-k\frac{r^2}{z_{0}}\right)}}{2\omega z_{0}}  \text{.}
\end{align}
As we can see, $\mathbf{s}$ is a purely transverse, azimuthally oriented vector field, since the longitudinal component $s^{z}$ is zero. For illustration, we depict the TSD, \mbox{$\mathbf{s}^{\bot}=\mathbf{s}_{H}^{\bot}+\mathbf{s}_{E}^{\bot}$} (see yellow arrowheads), on top of the Gaussian distribution of the $x$-component of the electric field intensity ($\left|E_{x}\right|^{2}$) in Fig.~\ref{fig:Gauss}(a).
Despite the cylindrical symmetry of the full TSD, its type (electric or magnetic) changes depending on the azimuth within the beam. To emphasize the spatially dependent composition of the TSD, we plot the individual contributions $s_{E}^{x}$, $s_{E}^{y}$, $s_{H}^{x}$, and $s_{H}^{y}$ in Fig.~\ref{fig:Gauss}(b). The TSD of the electric field exhibits a two-lobe pattern along the $x$-axis for $s_{E}^{y}$ ($s_{E}^{x}=0$), while the two lobes of the TSD of the magnetic field are arranged along the $y$-axis for $s_{H}^{x}$ ($s_{H}^{y}=0$). Different color codes are used to highlight the differences between magnetic and electric TSDs. The two distributions of $\mathbf{s}_{H}^{\bot}$ and $\mathbf{s}_{E}^{\bot}$ are rotated by $90^{\circ}$ with respect to each other, which is a direct consequence of the orthogonality of the magnetic and electric transverse field components of the linearly polarized Gaussian beam studied here exemplarily. 

In conclusion, the results we derived from this simplified model beam highlight the importance of considering the distributions of $\mathbf{s}_{H}^{\bot}$ and $\mathbf{s}_{E}^{\bot}$ individually and to distinguish between both quantities experimentally. While techniques for measuring $\mathbf{s}_{E}^{\bot}$ have been presented recently~\cite{OConnor2014,Neugebauer2015}, in the following we discuss a measurement concept for $\mathbf{s}_{H}^{\bot}$, allowing for a direct comparison between $\mathbf{s}_{H}^{\bot}$ and $\mathbf{s}_{E}^{\bot}$ in complex and highly confined light fields.

\section{Experimental concept}
\begin{figure*} 
  \includegraphics[width=0.85\textwidth]{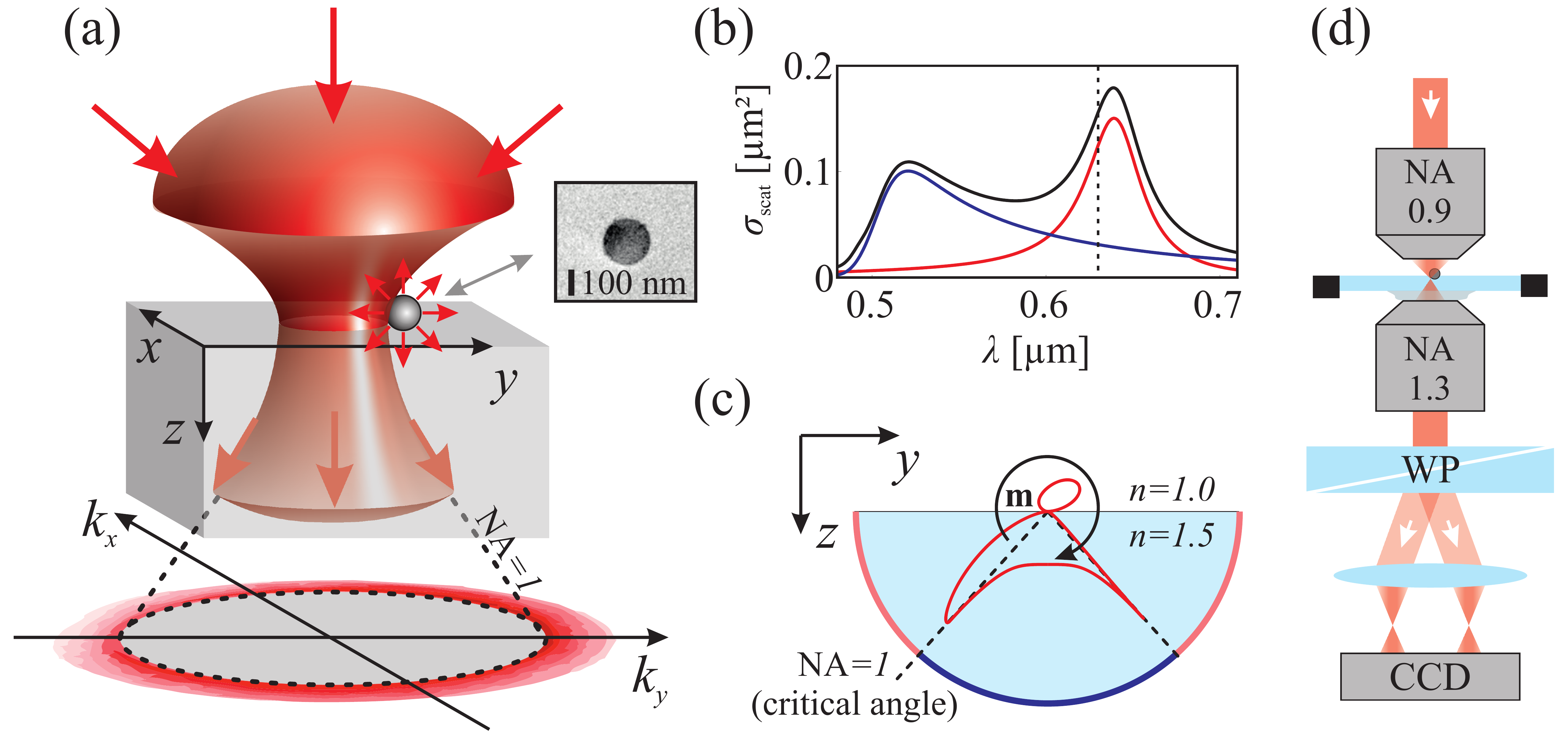}
  \caption{Experimental concept and setup. (a) Sketch of an incoming tightly focused beam being probed by a silicon nano-particle sitting on a glass substrate. The light scattered and transmitted into the glass half-space is collected in the far field. Only the light emitted into the solid angle above the critical angle ($\operatorname{NA}=1$, indicated by the dashed black lines) is required. Blocking the light collected below $\operatorname{NA}=1$ results in a ring-like far-field pattern visualized in red color coding. A scanning electron micrograph depicted as inset shows the particle (black scale bar indicates $100$~nm).	(b) Scattering cross-section of a core-shell nano-sphere (silicon core with a diameter of $158$~nm and silicon dioxide shell with $8$~nm thickness) calculated using Mie-theory. (c) Directional emission (red line) of a transversely spinning magnetic dipole (dipole moment spinning clockwise, $\mathbf{m}\propto \left(0,1,\imath\right)$, indicated by the black arrow) sitting on a glass substrate with refractive index $n=1.5$. (d) Experimental setup. An incoming monochromatic beam of light ($\lambda=630$~nm) is focused onto the nano-sphere sitting on a glass substrate by a microscope objective with a $\operatorname{NA}$ of 0.9. The transmitted light is collected by a microscope objective with an $\operatorname{NA}$ of 1.3. A Wollaston prism (WP) is utilized to determine the polarization state ($x$- and $y$-polarization) of the far-field pattern, which is measured by imaging the back focal plane of the lower microscope objective onto a CCD camera.}
  \label{fig:Setup}
\end{figure*}

Our experimental approach for simultaneously measuring $\mathbf{s}_{H}^{\bot}$ and $\mathbf{s}_{E}^{\bot}$ relies on a field probe that exhibits a magnetic as well as an electric dipole resonance. In this regard, suitable field probes, which support both types of modes are high-refractive-index nano-particles~\cite{Evlyukhin2012,Kuznetsov2012,Krasnok2012}. Here, we utilize a silicon ($\text{Si}$) nano-sphere with core radius $r_{\text{Si}}=79$~nm and an estimated silicon dioxide ($\text{SiO}_{2}$) shell of thickness $s=8$~nm as probe. The particle is sitting on a glass substrate [see sketch in Fig.~\ref{fig:Setup}(a)] attached to a 3D-piezo stage, enabling us to scan the field probe through the focal plane of a tightly focused beam. 
A scanning electron micrograph of the particle is shown as inset. In order to understand the actual scattering behavior of our probe, we first analyze its scattering cross-section [see Fig.~\ref{fig:Setup}(b)] using Mie-theory~\cite{Mie1908}. For the calculation, we assumed a particle in free space, not considering the glass substrate. The black line indicates the total scattering cross-section of the particle, with the red and blue lines representing the contributions of the magnetic and electric dipolar modes, respectively. Due to the spectral overlap between the electric and magnetic resonances, a generic input field excites a dipolar mode with simultaneously electric as well as magnetic contributions in the shown spectral range~\cite{Wozniak2015,Neugebauer2016,Evlyukhin2012,Kuznetsov2012}. With the objective to induce electric and magnetic dipole moments with a comparatively high efficiency, we choose an excitation wavelength between the maxima of both resonances (here $\lambda=630$~nm) for the TSD sensing experiment. Using a point dipole approximation, the electric ($\mathbf{p}$) and magnetic ($\mathbf{m}$) dipole moments of the particle are thus both proportional to the local excitation fields, $\mathbf{p}\propto\mathbf{E}$ and $\mathbf{m}\propto\mathbf{H}$, while higher order multipoles can be neglected. 

The direct link between the excitation field and the induced dipole moments is the basis of our TSD reconstruction approach. When we can determine the magnetic and the electric transversely spinning dipole moments of our probe particle from the light it scatters into the far field, we effectively measure the TSD of the excitation field~\cite{Neugebauer2015}. In order to achieve an unambiguous reconstruction of $\mathbf{s}_{H}^{\bot}$ and $\mathbf{s}_{E}^{\bot}$, a detailed analysis of the simultaneous emission of magnetic and electric dipoles close to a dielectric interface is required.

In comparison to a dipole in free space, the far-field emission pattern of a dipole above a dielectric substrate is strongly altered by the air-glass interface~\cite{Lukosz1977b,Novotny2006}. Due to the dominant emission of a dipole into the higher-index material~\cite{Lukosz1977b}, we are specifically interested in the light transmitted into the glass half-space. To calculate the directional emission, we use a plane-wave decomposition with the transverse electric ($E_{s}$) and transverse magnetic ($E_{p}$) polarization states as basis. Following ref.~\cite{Novotny2006} the far field in the glass half-space, $\mathbf{E}_{f}=E_{p}\mathbf{e}_{p}+E_{s}\mathbf{e}_{s}$, of an arbitrarily polarized electromagnetic dipole can be written in compact form as
\begin{align}\label{eqn:ff}
\mathbf{E}_{f}\left(k_{x},k_{y}\right)\propto
C\hat{\mathbf{T}}\left(\hat{\mathbf{M}}\mathbf{p}+\hat{\mathbf{R}}\hat{\mathbf{M}}\mathbf{m}/c_{0}\right)
\text{,}
\end{align}
with $C=\left(k_{0}^{2}n^{2}-k_{\bot}^{2}\right)^{1/2}/k_{z}\cdot\exp{\left[\imath k_{z}d\right]}$. The transverse wave number is defined as $k_{\bot}=\left(k_{x}^2+k_{y}^2\right)^{1/2}$, while the longitudinal wave number can be calculated by \mbox{$k_{z}=\left(k_{0}^{2}-k_{\bot}^{2}\right)^{1/2}$}. The parameter $d$ represents the distance between the dipole and the interface, and $c_{0}$ refers to the vacuum speed of light. The matrix $\hat{\mathbf{T}}$ consists of the Fresnel transmission coefficients $t_{s}$ and $t_{p}$~\cite{Novotny2006},
\begin{align}
\hat{\mathbf{T}}=&
\left(\begin{matrix}
t_{p}&0\\
0&t_{s}
\end{matrix}\right)\text{,}
\end{align}
while $\hat{\mathbf{M}}$ is a rotation matrix, representing the overlap of the electric and the magnetic dipole moments with the field vectors of the plane waves of the angular spectrum~\cite{Courtois1996},
\begin{align}
\hat{\mathbf{M}}=&
\left(\begin{matrix}
\frac{k_{x}k_{z}}{k_{\bot}k_{0}} &\frac{k_{y}k_{z}}{k_{\bot}k_{0}}&-\frac{k_{\bot}}{k_{0}}\\
-\frac{k_{y}}{k_{\bot}} &\frac{k_{x}}{k_{\bot}}&0
\end{matrix}\right)\text{.}
\end{align}
To calculate the far field of the magnetic components of the dipole emitter, a second rotation matrix $\hat{\mathbf{R}}$ is introduced, which is required due to interchanging electric and magnetic field vectors~\cite{Jackson1999,Novotny2006}:
\begin{align}
\hat{\mathbf{R}}=
\left(\begin{matrix}
0&1\\
-1&0
\end{matrix}\right)
\text{.}
\end{align}
We utilize Eq.~\eqref{eqn:ff} to exemplarily calculate the far-field emission pattern ($I=I_{p}+I_{s}\propto \left|E_{p}\right|^{2}+\left|E_{s}\right|^{2}$) of a magnetic dipole spinning around an axis parallel to the air-glass interface. For $\mathbf{m}=\left(0,1,\imath\right)$, we obtain the emission pattern depicted as side-view plot in Fig.~\ref{fig:Setup}(c), where, for the sake of completeness, we show the emission into the air half-space as well. As distance between the dipole and the interface in the calculations, we used the radius of the particle, $r_{0}=r_{\text{Si}}+s=87$~nm. We see that similar to a transversely spinning electric dipole moment (see for example refs.~\cite{Neugebauer2014,LeKien2016}), the transversely spinning magnetic dipole moment results in a directional far-field emission into the angular region above the critical angle ($k_{\bot}>k_{0}$, $k_{z}=\left(k_{0}^{2}-k_{\bot}^{2}\right)^{1/2}=\imath \left|k_{z}\right|$). By assuming $\mathbf{m}\propto\mathbf{H}$, this links the TSD of the magnetic field to the far-field directionality.

In this context, the objective of the following theoretical discussion is the derivation of a quantitative connection between the TSD (magnetic and electric) and the directional emission pattern of the probe particle above the critical angle. For that purpose, we need to calculate the difference of the light scattered into opposite transverse directions for a general electromagnetic dipole~\cite{Neugebauer2015}. First, we consider the directionality along the $x$-direction, $k_{x}=\pm k_{\bot}$ and $k_{y}=0$, and above the critical angle, $k_{\bot}>k_{0}$. Calculating the difference between the light scattered in the positive and negative $x$-direction for both polarization states, $\Delta_{x}^{k_{\bot}} I_{p}=I_{p}\left(k_{\bot},0\right)-I_{p}\left(-k_{\bot},0\right)$ and $\Delta_{x}^{k_{\bot}} I_{s}=I_{s}\left(k_{\bot},0\right)-I_{s}\left(-k_{\bot},0\right)$, results in
\begin{align}\label{eqn:DIpx}
\Delta_{x}^{k_{\bot}} I_{p}&= 
D\left|t_{p}\right|^{2}
\left[\frac{\left|k_{z}\right|\operatorname{Im}\left(p_{z}^{*}p_{x}\right)}{k_{0}} - 
\frac{\operatorname{Re}\left(m_{y}^{*}p_{z}\right)}{c_{0}}\right] \text{,}\\\label{eqn:DIsx}
\Delta_{x}^{k_{\bot}} I_{s}&= 
D\left|t_{s}\right|^{2}
\left[\frac{\left|k_{z}\right|\operatorname{Im}\left(m_{z}^{*}m_{x}\right)}{k_{0}c_{0}^{2}} + 
\frac{\operatorname{Re}\left(p_{y}^{*}m_{z}\right)}{c_{0}}\right]\text{,}
\end{align}
with $D=4 \left|C\right|^{2}k_{\bot}/k_{0}$. By performing a similar calculation for the $y$-direction, $\Delta_{y}^{k_{\bot}} I_{p}=I_{p}\left(0,-k_{\bot}\right)-I_{p}\left(0,k_{\bot}\right)$ and $\Delta_{y}^{k_{\bot}} I_{s}=I_{s}\left(0,-k_{\bot}\right)-I_{s}\left(0,k_{\bot}\right)$, we obtain
\begin{align}\label{eqn:DIpy}
\Delta_{y}^{k_{\bot}} I_{p}&= 
D\left|t_{p}\right|^{2}
\left[\frac{\left|k_{z}\right|\operatorname{Im}\left(p_{y}^{*}p_{z}\right)}{k_{0}} - 
\frac{\operatorname{Re}\left(p_{z}^{*}m_{x}\right)}{c_{0}}\right] \text{,}\\\label{eqn:DIsy}
\Delta_{y}^{k_{\bot}} I_{s}&= 
D\left|t_{s}\right|^{2}
\left[\frac{\left|k_{z}\right|\operatorname{Im}\left(m_{y}^{*}m_{z}\right)}{k_{0}c_{0}^{2}} + 
\frac{\operatorname{Re}\left(m_{z}^{*}p_{x}\right)}{c_{0}}\right]\text{.}
\end{align}
For each of the four Eqs.~\eqref{eqn:DIpx}-\eqref{eqn:DIsy}, we can discern two different terms. The first terms include only electric or magnetic dipole components, while the second terms consist of a mixture of both electric and magnetic dipole components. A comparison of the purely magnetic and purely electric terms with the magnetic and electric components of the TSD --- $s_{H}^{x}\propto \operatorname{Im}\left(H_{y}^{*}H_{z}\right)$, $s_{H}^{y}\propto \operatorname{Im}\left(H_{z}^{*}H_{x}\right)$, $s_{E}^{x}\propto \operatorname{Im}\left(E_{y}^{*}E_{z}\right)$, and $s_{E}^{y}\propto \operatorname{Im}\left(E_{z}^{*}E_{x}\right)$ --- reveals a strong similarity. Considering the aforementioned dipole approximation of the scattering response of the particle, $\mathbf{p}\propto\mathbf{E}$ and $\mathbf{m}\propto\mathbf{H}$, we see that the first terms in Eqs.~\eqref{eqn:DIpx}-\eqref{eqn:DIsy} are proportional to the individual components of the TSD. However, the four equations contain additional terms, which represent the interference of electric and magnetic dipole components. A simple difference measurement of the scattered light --- as it is discussed in ref.~\cite{Neugebauer2015} --- would therefore not be sufficient to reconstruct the TSD. Nonetheless, it is possible to unambiguously distinguish between the terms representing the TSD and the non-relevant electromagnetic interference terms by measuring the directional emission for two different transverse wave numbers, $k_{\bot 1}$ and $k_{\bot 2}$, since only the terms corresponding to the TSD exhibit factors depending on $k_{\bot}$. For example, by measuring $\Delta_{x}^{k_{\bot 1}} I_{s}$ and $\Delta_{x}^{k_{\bot 2}} I_{s}$ we result in two linearly independent equations, which can be solved for the term representing $s_{H}^{y}$. The same approach can be utilized for the three other components of the transverse spin density, $s_{H}^{x}$, $s_{E}^{x}$ \mbox{and $s_{E}^{y}$}.

With this theoretical consideration in mind, we can finally design an experimental procedure, capable of measuring the TSD of an incoming tightly focused beam. Figure~\ref{fig:Setup}(d) shows a sketch of our setup. A polarization tailored beam is tightly focused by a microscope objective with $\text{NA}=0.9$. The resulting focal field is probed by the $\text{Si}$-particle immobilized on a glass substrate. The probe can be scanned through the focal plane by a 3D-piezo stage. Below the substrate, an oil-immersion-type objective with $\text{NA}=1.3$ is collecting the light transmitted through the interface and scattered into the glass half-space. The far-field emission pattern of the particle, to be observed in the back focal plane (BFP) of the collection objective, is subsequently analyzed in its polarization distribution. Hence, the collected light is passed through a Wollaston prism (WP), splitting the beam into two orthogonal polarization states. Imaging the BFP with a lens through the WP onto a camera therefore results in two BFP images representing a decomposition into $x$- and $y$-polarization, respectively. 

\begin{figure} 
  \includegraphics[width=0.48\textwidth]{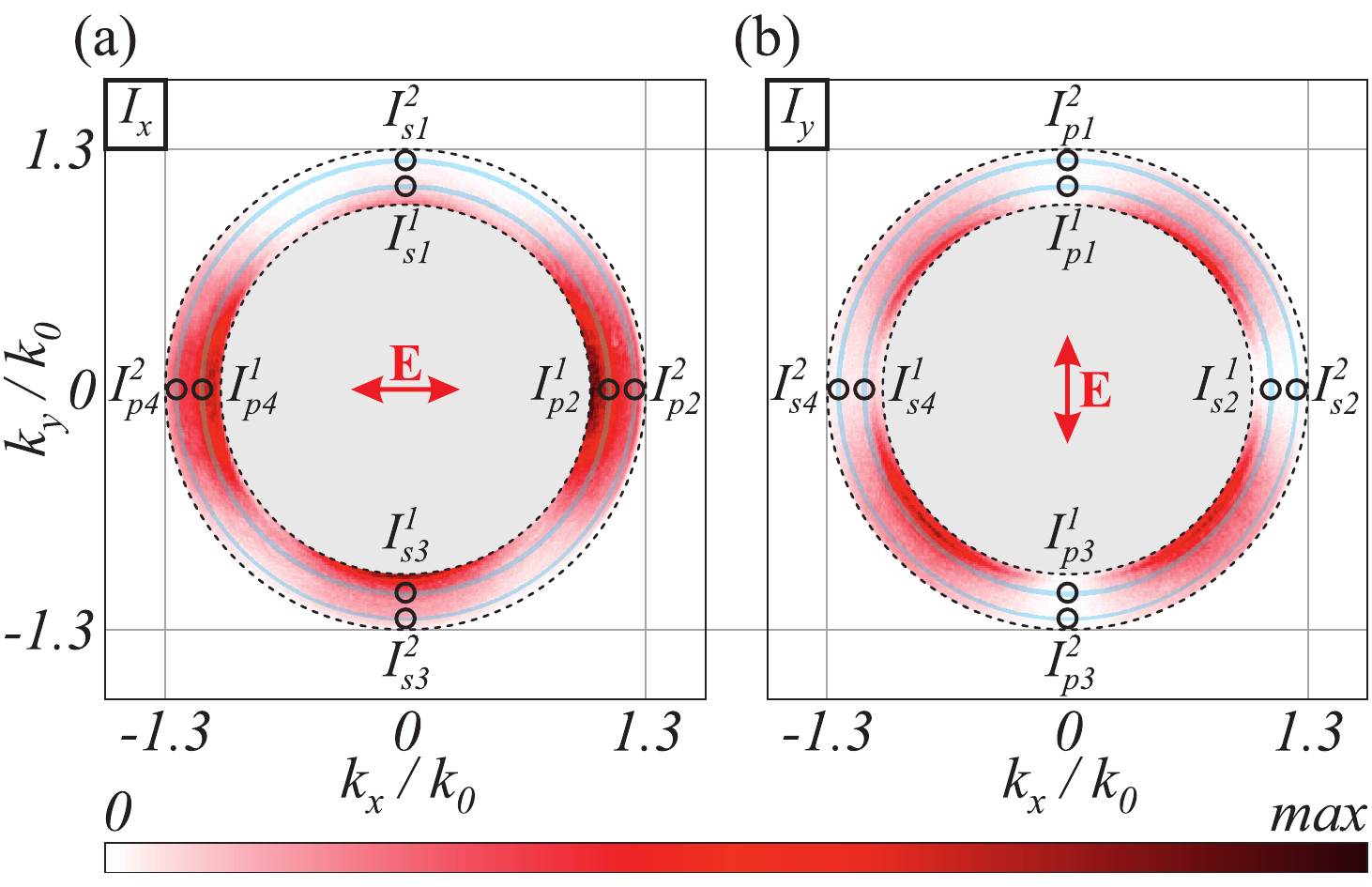}
  \caption{Polarization-resolved back focal plane (BFP) images. (a) and (b) show exemplarily measured $x$- and $y$-polarized BFP images in the angular range defined by $1 \leq k_{\bot}/k_{0}\leq 1.3$. Both images are normalized to their common maximum value. The inner dashed black circle corresponds to the critical angle $k_{\bot}/k_{0}=1$. The inner and outer semi transparent blue circles indicate $k_{\bot 1}/k_{0}\equiv 1.1$ and $k_{\bot 2}/k_{0} \equiv 1.25$. The outer dashed black circle indicates $k_{\bot}/k_{0}=1.3$, representing the $\text{NA}$ of the collection objective. Additional 8 small black circles mark regions in the BFP, for which an averaged intensity value is determined, $I_{si}^{j}$ and $I_{pi}^{j}$ with $i=1,2,3,4$ indicating the azimuthal position and $j=1,2$ referring to $k_{\bot 1}$ and $k_{\bot 2}$.}
  \label{fig:BFP}
\end{figure}

To exemplarily demonstrate the reconstruction of $\mathbf{s}_{E}^{\bot}$ and $\mathbf{s}_{H}^{\bot}$ from such polarization-resolved BFP images, we placed the $\text{Si}$-probe in the focal plane of a tightly focused linearly $x$-polarized Gaussian beam and shifted the particle with respect to the center of the focal spot by $150$~nm along the $y$-direction. The resulting $x$- and $y$-polarized BFP intensity distributions, $I_{x}$ and $I_{y}$, are shown in Figs~\ref{fig:BFP}(a)~and~(b).
In order to determine $\mathbf{s}_{E}^{\bot}$ and $\mathbf{s}_{H}^{\bot}$ for this position of the probe particle, we average the far-field intensity in $2\times4$ small regions in both BFP images (see small black circles) and obtain $I_{si}^{j}$ and $I_{pi}^{j}$ with $i=1,2,3,4$ indicating the azimuthal position and $j=1,2$ referring to two different transverse $k$-vectors $k_{\bot 1}/k_{0}\equiv 1.1$ and $k_{\bot 2}/k_{0} \equiv 1.25$. It is important to note that, although we measured the BFP images in the $x$- and $y$-polarization basis, we can assign the indices $p$ and $s$ to the averaged intensity values, since along the $k_{x}$- and $k_{y}$-axes in $k$-space, the transverse magnetic and transverse electric polarization basis coincides with the $x$- and $y$-polarization basis. Therefore, the distribution of $s_{H}^{x}\propto \operatorname{Im}\left(m_{y}^{*}m_{z}\right)$ can, for example, be calculated from $\Delta_{y}^{k_{\bot 1}} I_{s}=I_{s3}^{1}-I_{s1}^{1}$ and $\Delta_{y}^{k_{\bot 2}} I_{s}=I_{s3}^{2}-I_{s1}^{2}$. Correspondingly, we obtain $s_{E}^{x}$ from $\Delta_{y}^{k_{\bot 1}} I_{p}=I_{p3}^{1}-I_{p1}^{1}$ and $\Delta_{y}^{k_{\bot 2}} I_{p}=I_{p3}^{2}-I_{p1}^{2}$, $s_{H}^{y}$ from $\Delta_{x}^{k_{\bot 1}} I_{s}=I_{s2}^{1}-I_{s4}^{1}$ and $\Delta_{x}^{k_{\bot 2}} I_{s}=I_{s2}^{2}-I_{s4}^{2}$, and $s_{E}^{y}$ from $\Delta_{x}^{k_{\bot 1}} I_{p}=I_{p2}^{1}-I_{p4}^{1}$ and $\Delta_{x}^{k_{\bot 2}} I_{p}=I_{p2}^{2}-I_{p4}^{2}$. The actual measurement results, which represent scans of the particle through different tightly focused beams, are shown in the following.

\section{Experimental results and discussion}
\begin{figure*} 
  \includegraphics[width=1\textwidth]{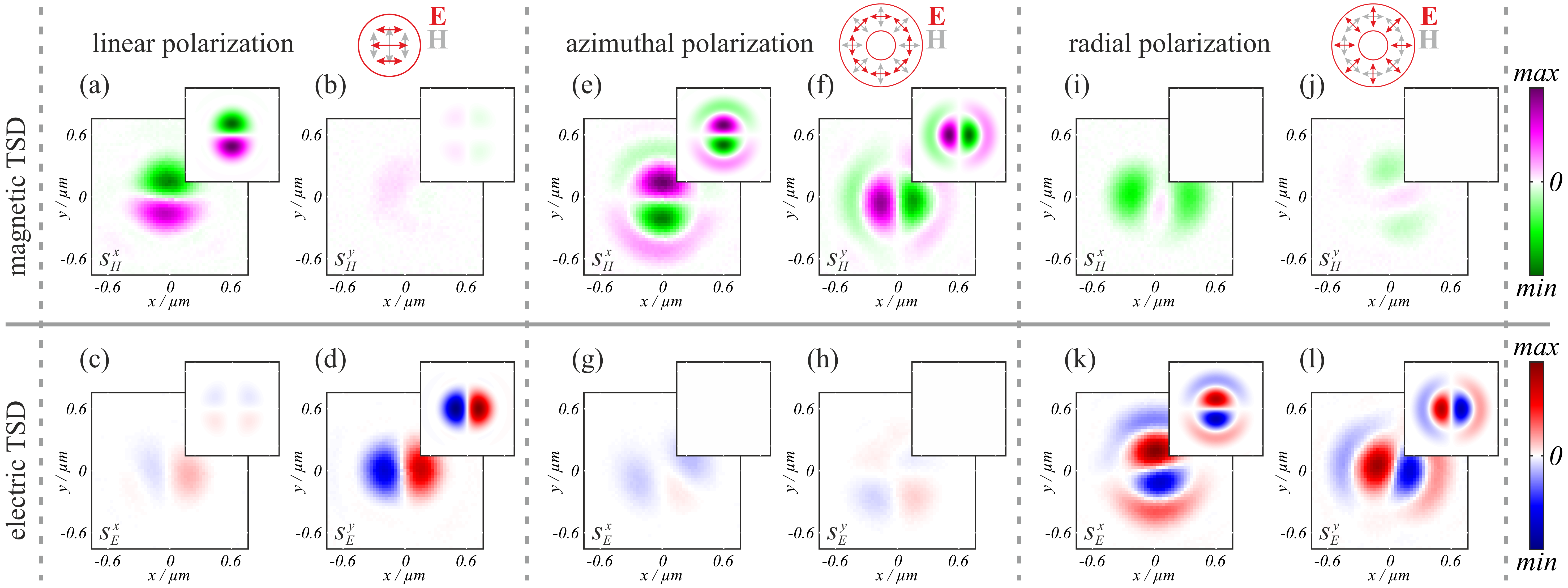}
  \caption{Experimentally measured and theoretically calculated focal distributions of the TSD of tightly focused linearly, azimuthally and radially polarized beams. The left column shows the TSD distributions of a tightly focused linearly polarized beam. (a) and (b) depict the $x$- and $y$-component of $\mathbf{s}_{H}$ while (c) and (d) depict the $x$- and $y$-component of $\mathbf{s}_{E}$. (e)-(h) display the corresponding distributions of a tightly focused azimuthally polarized beam (central column), and (i)-(l) present the corresponding distributions of a tightly focused radially polarized beam (right column). All distributions of the magnetic and the electric TSD are normalized to their common maximum value, respectively, in order to enable a direct comparison between all three beams.}
  \label{fig:LinearRadialAzimuthal}
\end{figure*}

At first, we utilize our approach to reconstruct the TSD components of a tightly focused linearly $x$-polarized Gaussian beam. Considering the simplified TSD distribution described in Eq.~\eqref{eq:Gauss_s} and depicted in Fig.~\ref{fig:Gauss}(a)~and~(b), we expect to obtain two-lobe patterns for $s_{H}^{x}$ and $s_{E}^{y}$ rotated by $90^{\circ}$ with respect to each other. The actual experimental results are shown in the left column of Fig.~\ref{fig:LinearRadialAzimuthal} while a sketch of the cross-section of the input beam is shown as inset above (red and gray vectors indicate $x$-polarized electric and $y$-polarized magnetic fields, respectively).
We clearly recognize the expected two-lobe patterns of $s_{H}^{x}$ and $s_{E}^{y}$ in Figs.~\ref{fig:LinearRadialAzimuthal}(a)~and~(d), and we observe their rotation of $90^{\circ}$ with respect to each other. The  experimental results are in very good agreement with the theoretical distributions (see insets) calculated with vectorial diffraction theory~\cite{Richards1959,Novotny2006}. Minor deviations are caused by imperfections of the incoming beam, the probe particle and the elements in the detection path. In particular, imaging the BFP after passing through the WP can be identified as one of the main sources of error, since the two partial beams are impinging onto the imaging lens under an angle [see Fig.~\ref{fig:Setup}(d)] and the optical path lengths through the WP are slightly different for $x$- and $y$-polarized light. It should also be mentioned here that in contrast to the simplified TSD distributions in Fig.~\ref{fig:Gauss}(b), where $s_{H}^{y}$ and $s_{E}^{x}$ are exactly zero, both distributions exhibit weak four-lobe patterns in the case of a tightly focused beam [see insets in Figs.~\ref{fig:LinearRadialAzimuthal}(b)~and~(c)]. Although the measured distributions of $s_{H}^{y}$ and $s_{E}^{x}$ are indeed much weaker than the two-lobe patterns of $s_{H}^{x}$ and $s_{E}^{y}$, they do not perfectly resemble the theoretical expectations. Nonetheless it can be seen that our measurement approach is sensitive and allows for demonstrating the main features of $\mathbf{s}^{\bot}_{E}$ and $\mathbf{s}^{\bot}_{H}$ for the case of a tightly focused linearly polarized Gaussian beam.

In order to verify our experimental scheme and additionally explore and highlight differences between the magnetic and the electric TSD in more complex light fields, we investigate two tightly focused cylindrical vector beams with azimuthal and radial polarization distributions. We have chosen these beams, because they can be transformed from one to the other by interchanging electric and magnetic fields, allowing for cross-checking of our experimental results. As an illustration, we depict sketches of the incoming azimuthally and radially polarized beams as insets on top of the central and right columns of Fig.~\ref{fig:LinearRadialAzimuthal}, respectively. As we can see, the azimuthal polarization of the electric field is accompanied by a radially polarized magnetic field while the radial polarization of the electric field implies an azimuthally polarized magnetic field. 

An important feature of azimuthally polarized beams is their purely transverse electric field distribution, which remains purely transverse even when the beam is tightly focused~\cite{Youngworth2000}. Therefore, the electric TSD must be zero. In contrast, the magnetic field of such a tightly focused beam exhibits a strong longitudinal component~\cite{Wozniak2015} and, in particular, transversely spinning magnetic fields. The theoretical predictions and the experimentally measured distributions of $\mathbf{s}_{H}^{\bot}$ and $\mathbf{s}_{E}^{\bot}$ are shown in Fig.~\ref{fig:LinearRadialAzimuthal}(e)-(h). We see a good overlap of theory and experiment, effectively verifying the aforementioned statements. In particular, we see a strong magnetic TSD and a very weak (theoretically zero valued) electric TSD.

When comparing these results with the distributions of the tightly focused radially polarized beam plotted in Fig.~\ref{fig:LinearRadialAzimuthal}(i)-(l), we see that, as expected, $\mathbf{s}_{H}^{\bot}$ and $\mathbf{s}_{E}^{\bot}$ are essentially interchanged. This time, we obtain a strong electric TSD accompanied by a weak (ideally zero valued) magnetic TSD. The apparent minor rotations of the experimental distributions of $\mathbf{s}_{E}^{\bot}$ with respect to the theoretical prediction are caused by aberrations of the incoming beam and deviations in the response of the probe particle in combination with the aforementioned imperfections of the detection path.

\section{Conclusion}
In conclusion, we presented a probe-based scanning technique which allows for the simultaneous reconstruction of the TSD of the magnetic and the electric field. This was achieved by analyzing the far-field directionality of the light scattered off the nano-probe. We utilized the technique to emphasize the importance of distinguishing the magnetic and the electric components of the TSD in the case of highly confined light. In the process, we demonstrated the difference between the distributions of $\mathbf{s}_{H}^{\bot}$ and $\mathbf{s}_{E}^{\bot}$ in a tightly focused linearly polarized beam. In particular, we showed their $90^{\circ}$ rotation with respect to each other. Additionally, we investigated $\mathbf{s}_{H}^{\bot}$ and $\mathbf{s}_{E}^{\bot}$ in tightly focused azimuthally and radially polarized beams. Thereby, we highlighted that the radially polarized beam exhibits purely electric TSD, while the azimuthally polarized beam exhibits purely magnetic TSD. This implies, that these beams can be used in experiments trying to distinguish, whether an effect depends on $\mathbf{s}_{H}^{\bot}$, $\mathbf{s}_{E}^{\bot}$ or their interplay. 

From a general and more conceptional point of view, our results demonstrate the relevance of the dual symmetry (electromagnetic democracy) representation of the spin angular momentum of light. In this regard, our manuscript takes its place alongside recent experimental efforts to distinguish the different components of linear and angular momentum of light~\cite{Antognozzi2016}.

Finally, by being able to influence and tailor the magnetic and electric part of the TSD separately, we anticipate that the spin-momentum locking of transversely spinning magnetic dipoles, an effect which has been recently investigated in refs.~\cite{MichelaF.PicardiAlejandroManjavacasAnatolyV.Zayats2017,Wang2017}, will gain relevance in upcoming experimental and theoretical studies, similar to the spin-momentum locking of spinning electric dipoles~\cite{Aiello2015,Bliokh2015,Lodahl2017}.
\begin{acknowledgments}
We gratefully acknowledge fruitful discussions with Andrea Aiello and Gerd Leuchs.
\end{acknowledgments}
%
\bibliography{bib}

\begin{thebibliography}{42}%
\makeatletter
\providecommand \@ifxundefined [1]{%
 \@ifx{#1\undefined}
}%
\providecommand \@ifnum [1]{%
 \ifnum #1\expandafter \@firstoftwo
 \else \expandafter \@secondoftwo
 \fi
}%
\providecommand \@ifx [1]{%
 \ifx #1\expandafter \@firstoftwo
 \else \expandafter \@secondoftwo
 \fi
}%
\providecommand \natexlab [1]{#1}%
\providecommand \enquote  [1]{``#1''}%
\providecommand \bibnamefont  [1]{#1}%
\providecommand \bibfnamefont [1]{#1}%
\providecommand \citenamefont [1]{#1}%
\providecommand \href@noop [0]{\@secondoftwo}%
\providecommand \href [0]{\begingroup \@sanitize@url \@href}%
\providecommand \@href[1]{\@@startlink{#1}\@@href}%
\providecommand \@@href[1]{\endgroup#1\@@endlink}%
\providecommand \@sanitize@url [0]{\catcode `\\12\catcode `\$12\catcode
  `\&12\catcode `\#12\catcode `\^12\catcode `\_12\catcode `\%12\relax}%
\providecommand \@@startlink[1]{}%
\providecommand \@@endlink[0]{}%
\providecommand \url  [0]{\begingroup\@sanitize@url \@url }%
\providecommand \@url [1]{\endgroup\@href {#1}{\urlprefix }}%
\providecommand \urlprefix  [0]{URL }%
\providecommand \Eprint [0]{\href }%
\providecommand \doibase [0]{http://dx.doi.org/}%
\providecommand \selectlanguage [0]{\@gobble}%
\providecommand \bibinfo  [0]{\@secondoftwo}%
\providecommand \bibfield  [0]{\@secondoftwo}%
\providecommand \translation [1]{[#1]}%
\providecommand \BibitemOpen [0]{}%
\providecommand \bibitemStop [0]{}%
\providecommand \bibitemNoStop [0]{.\EOS\space}%
\providecommand \EOS [0]{\spacefactor3000\relax}%
\providecommand \BibitemShut  [1]{\csname bibitem#1\endcsname}%
\let\auto@bib@innerbib\@empty
\bibitem [{\citenamefont {Aiello}\ \emph {et~al.}(2015)\citenamefont {Aiello},
  \citenamefont {Banzer}, \citenamefont {Neugebauer},\ and\ \citenamefont
  {Leuchs}}]{Aiello2015}%
  \BibitemOpen
  \bibfield  {author} {\bibinfo {author} {\bibfnamefont {A.}~\bibnamefont
  {Aiello}}, \bibinfo {author} {\bibfnamefont {P.}~\bibnamefont {Banzer}},
  \bibinfo {author} {\bibfnamefont {M.}~\bibnamefont {Neugebauer}}, \ and\
  \bibinfo {author} {\bibfnamefont {G.}~\bibnamefont {Leuchs}},\ }\href
  {\doibase 10.1038/nphoton.2015.203} {\bibfield  {journal} {\bibinfo
  {journal} {Nat. Photon.}\ }\textbf {\bibinfo {volume} {9}},\ \bibinfo {pages}
  {789} (\bibinfo {year} {2015})}\BibitemShut {NoStop}%
\bibitem [{\citenamefont {Bliokh}\ \emph
  {et~al.}(2015{\natexlab{a}})\citenamefont {Bliokh}, \citenamefont
  {Rodr{\'{i}}guez-Fortu{\~{n}}o}, \citenamefont {Nori},\ and\ \citenamefont
  {Zayats}}]{Bliokh2015}%
  \BibitemOpen
  \bibfield  {author} {\bibinfo {author} {\bibfnamefont {K.}~\bibnamefont
  {Bliokh}}, \bibinfo {author} {\bibfnamefont {F.}~\bibnamefont
  {Rodr{\'{i}}guez-Fortu{\~{n}}o}}, \bibinfo {author} {\bibfnamefont
  {F.}~\bibnamefont {Nori}}, \ and\ \bibinfo {author} {\bibfnamefont
  {A.}~\bibnamefont {Zayats}},\ }\href {\doibase 10.1038/nphoton.2015.201}
  {\bibfield  {journal} {\bibinfo  {journal} {Nat. Photon.}\ }\textbf {\bibinfo
  {volume} {9}},\ \bibinfo {pages} {796} (\bibinfo {year}
  {2015}{\natexlab{a}})}\BibitemShut {NoStop}%
\bibitem [{\citenamefont {Petersen}\ \emph {et~al.}(2014)\citenamefont
  {Petersen}, \citenamefont {Volz},\ and\ \citenamefont
  {Rauschenbeutel}}]{Petersen2014}%
  \BibitemOpen
  \bibfield  {author} {\bibinfo {author} {\bibfnamefont {J.}~\bibnamefont
  {Petersen}}, \bibinfo {author} {\bibfnamefont {J.}~\bibnamefont {Volz}}, \
  and\ \bibinfo {author} {\bibfnamefont {A.}~\bibnamefont {Rauschenbeutel}},\
  }\href {\doibase 10.1126/science.1257671} {\bibfield  {journal} {\bibinfo
  {journal} {Science}\ }\textbf {\bibinfo {volume} {346}},\ \bibinfo {pages}
  {67} (\bibinfo {year} {2014})}\BibitemShut {NoStop}%
\bibitem [{\citenamefont {S{\"{o}}llner}\ \emph {et~al.}(2015)\citenamefont
  {S{\"{o}}llner}, \citenamefont {Mahmoodian}, \citenamefont {Hansen},
  \citenamefont {Midolo}, \citenamefont {Javadi}, \citenamefont
  {Kir{\v{s}}anskė}, \citenamefont {Pregnolato}, \citenamefont {El-Ella},
  \citenamefont {Lee}, \citenamefont {Song}, \citenamefont {Stobbe},\ and\
  \citenamefont {Lodahl}}]{Sollner2015}%
  \BibitemOpen
  \bibfield  {author} {\bibinfo {author} {\bibfnamefont {I.}~\bibnamefont
  {S{\"{o}}llner}}, \bibinfo {author} {\bibfnamefont {S.}~\bibnamefont
  {Mahmoodian}}, \bibinfo {author} {\bibfnamefont {S.~L.}\ \bibnamefont
  {Hansen}}, \bibinfo {author} {\bibfnamefont {L.}~\bibnamefont {Midolo}},
  \bibinfo {author} {\bibfnamefont {A.}~\bibnamefont {Javadi}}, \bibinfo
  {author} {\bibfnamefont {G.}~\bibnamefont {Kir{\v{s}}anskė}}, \bibinfo
  {author} {\bibfnamefont {T.}~\bibnamefont {Pregnolato}}, \bibinfo {author}
  {\bibfnamefont {H.}~\bibnamefont {El-Ella}}, \bibinfo {author} {\bibfnamefont
  {E.~H.}\ \bibnamefont {Lee}}, \bibinfo {author} {\bibfnamefont {J.~D.}\
  \bibnamefont {Song}}, \bibinfo {author} {\bibfnamefont {S.}~\bibnamefont
  {Stobbe}}, \ and\ \bibinfo {author} {\bibfnamefont {P.}~\bibnamefont
  {Lodahl}},\ }\href {\doibase 10.1038/nnano.2015.159} {\bibfield  {journal}
  {\bibinfo  {journal} {Nat. Nanotechnol.}\ } (\bibinfo {year} {2015}),\
  10.1038/nnano.2015.159}\BibitemShut {NoStop}%
\bibitem [{\citenamefont {Young}\ \emph {et~al.}(2015)\citenamefont {Young},
  \citenamefont {Thijssen}, \citenamefont {Beggs}, \citenamefont {Kuipers},
  \citenamefont {Rarity},\ and\ \citenamefont {Oulton}}]{Young2015}%
  \BibitemOpen
  \bibfield  {author} {\bibinfo {author} {\bibfnamefont {A.~B.}\ \bibnamefont
  {Young}}, \bibinfo {author} {\bibfnamefont {A.}~\bibnamefont {Thijssen}},
  \bibinfo {author} {\bibfnamefont {D.~M.}\ \bibnamefont {Beggs}}, \bibinfo
  {author} {\bibfnamefont {L.}~\bibnamefont {Kuipers}}, \bibinfo {author}
  {\bibfnamefont {J.~G.}\ \bibnamefont {Rarity}}, \ and\ \bibinfo {author}
  {\bibfnamefont {R.}~\bibnamefont {Oulton}},\ }\href {\doibase
  10.1103/PhysRevLett.115.153901} {\bibfield  {journal} {\bibinfo  {journal}
  {Phys. Rev. Lett.}\ }\textbf {\bibinfo {volume} {115}},\ \bibinfo {pages}
  {153901} (\bibinfo {year} {2015})}\BibitemShut {NoStop}%
\bibitem [{\citenamefont {Lodahl}\ \emph {et~al.}(2017)\citenamefont {Lodahl},
  \citenamefont {Mahmoodian}, \citenamefont {Stobbe}, \citenamefont
  {Rauschenbeutel}, \citenamefont {Schneeweiss}, \citenamefont {Volz},
  \citenamefont {Pichler},\ and\ \citenamefont {Zoller}}]{Lodahl2017}%
  \BibitemOpen
  \bibfield  {author} {\bibinfo {author} {\bibfnamefont {P.}~\bibnamefont
  {Lodahl}}, \bibinfo {author} {\bibfnamefont {S.}~\bibnamefont {Mahmoodian}},
  \bibinfo {author} {\bibfnamefont {S.}~\bibnamefont {Stobbe}}, \bibinfo
  {author} {\bibfnamefont {A.}~\bibnamefont {Rauschenbeutel}}, \bibinfo
  {author} {\bibfnamefont {P.}~\bibnamefont {Schneeweiss}}, \bibinfo {author}
  {\bibfnamefont {J.}~\bibnamefont {Volz}}, \bibinfo {author} {\bibfnamefont
  {H.}~\bibnamefont {Pichler}}, \ and\ \bibinfo {author} {\bibfnamefont
  {P.}~\bibnamefont {Zoller}},\ }\href {\doibase 10.1038/nature21037}
  {\bibfield  {journal} {\bibinfo  {journal} {Nature}\ }\textbf {\bibinfo
  {volume} {541}},\ \bibinfo {pages} {473} (\bibinfo {year}
  {2017})}\BibitemShut {NoStop}%
\bibitem [{\citenamefont {Bliokh}\ and\ \citenamefont
  {Nori}(2012)}]{Bliokh2012}%
  \BibitemOpen
  \bibfield  {author} {\bibinfo {author} {\bibfnamefont {K.~Y.}\ \bibnamefont
  {Bliokh}}\ and\ \bibinfo {author} {\bibfnamefont {F.}~\bibnamefont {Nori}},\
  }\href {\doibase 10.1103/PhysRevA.85.061801} {\bibfield  {journal} {\bibinfo
  {journal} {Phys. Rev. A}\ }\textbf {\bibinfo {volume} {85}},\ \bibinfo
  {pages} {061801} (\bibinfo {year} {2012})}\BibitemShut {NoStop}%
\bibitem [{\citenamefont {Kim}\ \emph {et~al.}(2012)\citenamefont {Kim},
  \citenamefont {Lee}, \citenamefont {Kim}, \citenamefont {Jung},\ and\
  \citenamefont {Lee}}]{Kim2012}%
  \BibitemOpen
  \bibfield  {author} {\bibinfo {author} {\bibfnamefont {K.~Y.}\ \bibnamefont
  {Kim}}, \bibinfo {author} {\bibfnamefont {I.~M.}\ \bibnamefont {Lee}},
  \bibinfo {author} {\bibfnamefont {J.}~\bibnamefont {Kim}}, \bibinfo {author}
  {\bibfnamefont {J.}~\bibnamefont {Jung}}, \ and\ \bibinfo {author}
  {\bibfnamefont {B.}~\bibnamefont {Lee}},\ }\href {\doibase
  10.1103/PhysRevA.86.063805} {\bibfield  {journal} {\bibinfo  {journal} {Phys.
  Rev. A}\ }\textbf {\bibinfo {volume} {86}},\ \bibinfo {pages} {1} (\bibinfo
  {year} {2012})}\BibitemShut {NoStop}%
\bibitem [{\citenamefont {Canaguier-Durand}\ and\ \citenamefont
  {Genet}(2014)}]{Canaguier-Durand2014}%
  \BibitemOpen
  \bibfield  {author} {\bibinfo {author} {\bibfnamefont {A.}~\bibnamefont
  {Canaguier-Durand}}\ and\ \bibinfo {author} {\bibfnamefont {C.}~\bibnamefont
  {Genet}},\ }\href {\doibase 10.1103/PhysRevA.89.033841} {\bibfield  {journal}
  {\bibinfo  {journal} {Phys. Rev. A}\ }\textbf {\bibinfo {volume} {89}},\
  \bibinfo {pages} {033841} (\bibinfo {year} {2014})}\BibitemShut {NoStop}%
\bibitem [{\citenamefont {Saha}\ \emph {et~al.}(2016)\citenamefont {Saha},
  \citenamefont {Singh}, \citenamefont {Ray}, \citenamefont {Banerjee},
  \citenamefont {Gupta},\ and\ \citenamefont {Ghosh}}]{Saha2016}%
  \BibitemOpen
  \bibfield  {author} {\bibinfo {author} {\bibfnamefont {S.}~\bibnamefont
  {Saha}}, \bibinfo {author} {\bibfnamefont {A.~K.}\ \bibnamefont {Singh}},
  \bibinfo {author} {\bibfnamefont {S.~K.}\ \bibnamefont {Ray}}, \bibinfo
  {author} {\bibfnamefont {A.}~\bibnamefont {Banerjee}}, \bibinfo {author}
  {\bibfnamefont {S.~D.}\ \bibnamefont {Gupta}}, \ and\ \bibinfo {author}
  {\bibfnamefont {N.}~\bibnamefont {Ghosh}},\ }\href {\doibase
  10.1364/OL.41.004499} {\bibfield  {journal} {\bibinfo  {journal} {Opt.
  Lett.}\ }\textbf {\bibinfo {volume} {41}},\ \bibinfo {pages} {4499} (\bibinfo
  {year} {2016})}\BibitemShut {NoStop}%
\bibitem [{\citenamefont {Junge}\ \emph {et~al.}(2013)\citenamefont {Junge},
  \citenamefont {O'Shea}, \citenamefont {Volz},\ and\ \citenamefont
  {Rauschenbeutel}}]{Junge2013}%
  \BibitemOpen
  \bibfield  {author} {\bibinfo {author} {\bibfnamefont {C.}~\bibnamefont
  {Junge}}, \bibinfo {author} {\bibfnamefont {D.}~\bibnamefont {O'Shea}},
  \bibinfo {author} {\bibfnamefont {J.}~\bibnamefont {Volz}}, \ and\ \bibinfo
  {author} {\bibfnamefont {A.}~\bibnamefont {Rauschenbeutel}},\ }\href
  {\doibase 10.1103/PhysRevLett.110.213604} {\bibfield  {journal} {\bibinfo
  {journal} {Phys. Rev. Lett.}\ }\textbf {\bibinfo {volume} {110}},\ \bibinfo
  {pages} {213604} (\bibinfo {year} {2013})}\BibitemShut {NoStop}%
\bibitem [{\citenamefont {Banzer}\ \emph {et~al.}(2013)\citenamefont {Banzer},
  \citenamefont {Neugebauer}, \citenamefont {Aiello}, \citenamefont
  {Marquardt}, \citenamefont {Lindlein}, \citenamefont {Bauer},\ and\
  \citenamefont {Leuchs}}]{Banzer2013}%
  \BibitemOpen
  \bibfield  {author} {\bibinfo {author} {\bibfnamefont {P.}~\bibnamefont
  {Banzer}}, \bibinfo {author} {\bibfnamefont {M.}~\bibnamefont {Neugebauer}},
  \bibinfo {author} {\bibfnamefont {A.}~\bibnamefont {Aiello}}, \bibinfo
  {author} {\bibfnamefont {C.}~\bibnamefont {Marquardt}}, \bibinfo {author}
  {\bibfnamefont {N.}~\bibnamefont {Lindlein}}, \bibinfo {author}
  {\bibfnamefont {T.}~\bibnamefont {Bauer}}, \ and\ \bibinfo {author}
  {\bibfnamefont {G.}~\bibnamefont {Leuchs}},\ }\href {\doibase
  10.2971/jeos.2013.13032} {\bibfield  {journal} {\bibinfo  {journal} {J. Eur.
  Opt. Soc, Rapid Publ.}\ }\textbf {\bibinfo {volume} {8}},\ \bibinfo {pages}
  {13032} (\bibinfo {year} {2013})}\BibitemShut {NoStop}%
\bibitem [{\citenamefont {Neugebauer}\ \emph {et~al.}(2015)\citenamefont
  {Neugebauer}, \citenamefont {Bauer}, \citenamefont {Aiello},\ and\
  \citenamefont {Banzer}}]{Neugebauer2015}%
  \BibitemOpen
  \bibfield  {author} {\bibinfo {author} {\bibfnamefont {M.}~\bibnamefont
  {Neugebauer}}, \bibinfo {author} {\bibfnamefont {T.}~\bibnamefont {Bauer}},
  \bibinfo {author} {\bibfnamefont {A.}~\bibnamefont {Aiello}}, \ and\ \bibinfo
  {author} {\bibfnamefont {P.}~\bibnamefont {Banzer}},\ }\href {\doibase
  10.1103/PhysRevLett.114.063901} {\bibfield  {journal} {\bibinfo  {journal}
  {Phys. Rev. Lett.}\ }\textbf {\bibinfo {volume} {114}},\ \bibinfo {pages}
  {063901} (\bibinfo {year} {2015})}\BibitemShut {NoStop}%
\bibitem [{\citenamefont {Chen}\ \emph {et~al.}(2017)\citenamefont {Chen},
  \citenamefont {Wan}, \citenamefont {Kong},\ and\ \citenamefont
  {Zhan}}]{Chen2017}%
  \BibitemOpen
  \bibfield  {author} {\bibinfo {author} {\bibfnamefont {J.}~\bibnamefont
  {Chen}}, \bibinfo {author} {\bibfnamefont {C.}~\bibnamefont {Wan}}, \bibinfo
  {author} {\bibfnamefont {L.~J.}\ \bibnamefont {Kong}}, \ and\ \bibinfo
  {author} {\bibfnamefont {Q.}~\bibnamefont {Zhan}},\ }\href {\doibase
  10.1364/OE.25.019517} {\bibfield  {journal} {\bibinfo  {journal} {Optics
  Express}\ }\textbf {\bibinfo {volume} {25}},\ \bibinfo {pages} {19517}
  (\bibinfo {year} {2017})}\BibitemShut {NoStop}%
\bibitem [{\citenamefont {Rodr{\'{i}}guez-Fortu{\~{n}}o}\ \emph
  {et~al.}(2013)\citenamefont {Rodr{\'{i}}guez-Fortu{\~{n}}o}, \citenamefont
  {Marino}, \citenamefont {Ginzburg}, \citenamefont {O'Connor}, \citenamefont
  {Mart{\'{i}}nez}, \citenamefont {Wurtz},\ and\ \citenamefont
  {Zayats}}]{Rodriguez-Fortuno2013}%
  \BibitemOpen
  \bibfield  {author} {\bibinfo {author} {\bibfnamefont {F.~J.}\ \bibnamefont
  {Rodr{\'{i}}guez-Fortu{\~{n}}o}}, \bibinfo {author} {\bibfnamefont
  {G.}~\bibnamefont {Marino}}, \bibinfo {author} {\bibfnamefont
  {P.}~\bibnamefont {Ginzburg}}, \bibinfo {author} {\bibfnamefont
  {D.}~\bibnamefont {O'Connor}}, \bibinfo {author} {\bibfnamefont
  {A.}~\bibnamefont {Mart{\'{i}}nez}}, \bibinfo {author} {\bibfnamefont
  {G.~A.}\ \bibnamefont {Wurtz}}, \ and\ \bibinfo {author} {\bibfnamefont
  {A.~V.}\ \bibnamefont {Zayats}},\ }\href {\doibase 10.1126/science.1233739}
  {\bibfield  {journal} {\bibinfo  {journal} {Science}\ }\textbf {\bibinfo
  {volume} {340}},\ \bibinfo {pages} {328} (\bibinfo {year}
  {2013})}\BibitemShut {NoStop}%
\bibitem [{\citenamefont {{Van Mechelen}}\ and\ \citenamefont
  {Jacob}(2016)}]{VanMechelen2016}%
  \BibitemOpen
  \bibfield  {author} {\bibinfo {author} {\bibfnamefont {T.}~\bibnamefont {{Van
  Mechelen}}}\ and\ \bibinfo {author} {\bibfnamefont {Z.}~\bibnamefont
  {Jacob}},\ }\href {\doibase 10.1364/OPTICA.3.000118} {\bibfield  {journal}
  {\bibinfo  {journal} {Optica}\ }\textbf {\bibinfo {volume} {3}},\ \bibinfo
  {pages} {118} (\bibinfo {year} {2016})}\BibitemShut {NoStop}%
\bibitem [{\citenamefont {Bliokh}\ \emph
  {et~al.}(2015{\natexlab{b}})\citenamefont {Bliokh}, \citenamefont
  {Smirnova},\ and\ \citenamefont {Nori}}]{Bliokh2015q}%
  \BibitemOpen
  \bibfield  {author} {\bibinfo {author} {\bibfnamefont {K.~Y.}\ \bibnamefont
  {Bliokh}}, \bibinfo {author} {\bibfnamefont {D.}~\bibnamefont {Smirnova}}, \
  and\ \bibinfo {author} {\bibfnamefont {F.}~\bibnamefont {Nori}},\ }\href
  {\doibase 10.1126/science.aaa9519} {\bibfield  {journal} {\bibinfo  {journal}
  {Science}\ }\textbf {\bibinfo {volume} {348}},\ \bibinfo {pages} {1448}
  (\bibinfo {year} {2015}{\natexlab{b}})}\BibitemShut {NoStop}%
\bibitem [{\citenamefont {Neugebauer}\ \emph {et~al.}(2014)\citenamefont
  {Neugebauer}, \citenamefont {Bauer}, \citenamefont {Banzer},\ and\
  \citenamefont {Leuchs}}]{Neugebauer2014}%
  \BibitemOpen
  \bibfield  {author} {\bibinfo {author} {\bibfnamefont {M.}~\bibnamefont
  {Neugebauer}}, \bibinfo {author} {\bibfnamefont {T.}~\bibnamefont {Bauer}},
  \bibinfo {author} {\bibfnamefont {P.}~\bibnamefont {Banzer}}, \ and\ \bibinfo
  {author} {\bibfnamefont {G.}~\bibnamefont {Leuchs}},\ }\href
  {http://pubs.acs.org/doi/abs/10.1021/nl5003526} {\bibfield  {journal}
  {\bibinfo  {journal} {Nano Lett.}\ }\textbf {\bibinfo {volume} {14}},\
  \bibinfo {pages} {2546} (\bibinfo {year} {2014})}\BibitemShut {NoStop}%
\bibitem [{\citenamefont {le~Feber}\ \emph {et~al.}(2015)\citenamefont
  {le~Feber}, \citenamefont {Rotenberg},\ and\ \citenamefont
  {Kuipers}}]{LeFeber2015}%
  \BibitemOpen
  \bibfield  {author} {\bibinfo {author} {\bibfnamefont {B.}~\bibnamefont
  {le~Feber}}, \bibinfo {author} {\bibfnamefont {N.}~\bibnamefont {Rotenberg}},
  \ and\ \bibinfo {author} {\bibfnamefont {L.}~\bibnamefont {Kuipers}},\ }\href
  {\doibase 10.1038/ncomms7695} {\bibfield  {journal} {\bibinfo  {journal}
  {Nat. Commun.}\ }\textbf {\bibinfo {volume} {6}},\ \bibinfo {pages} {6695}
  (\bibinfo {year} {2015})}\BibitemShut {NoStop}%
\bibitem [{\citenamefont {Sayrin}\ \emph {et~al.}(2015)\citenamefont {Sayrin},
  \citenamefont {Junge}, \citenamefont {Mitsch}, \citenamefont {Albrecht},
  \citenamefont {O'Shea}, \citenamefont {Schneeweiss}, \citenamefont {Volz},\
  and\ \citenamefont {Rauschenbeutel}}]{Sayrin2015a}%
  \BibitemOpen
  \bibfield  {author} {\bibinfo {author} {\bibfnamefont {C.}~\bibnamefont
  {Sayrin}}, \bibinfo {author} {\bibfnamefont {C.}~\bibnamefont {Junge}},
  \bibinfo {author} {\bibfnamefont {R.}~\bibnamefont {Mitsch}}, \bibinfo
  {author} {\bibfnamefont {B.}~\bibnamefont {Albrecht}}, \bibinfo {author}
  {\bibfnamefont {D.}~\bibnamefont {O'Shea}}, \bibinfo {author} {\bibfnamefont
  {P.}~\bibnamefont {Schneeweiss}}, \bibinfo {author} {\bibfnamefont
  {J.}~\bibnamefont {Volz}}, \ and\ \bibinfo {author} {\bibfnamefont
  {A.}~\bibnamefont {Rauschenbeutel}},\ }\href {\doibase
  10.1103/PhysRevX.5.041036} {\bibfield  {journal} {\bibinfo  {journal} {Phys.
  Rev. X}\ }\textbf {\bibinfo {volume} {5}},\ \bibinfo {pages} {041036}
  (\bibinfo {year} {2015})}\BibitemShut {NoStop}%
\bibitem [{\citenamefont {Scheucher}\ \emph {et~al.}(2016)\citenamefont
  {Scheucher}, \citenamefont {Hilico}, \citenamefont {Will}, \citenamefont
  {Volz},\ and\ \citenamefont {Rauschenbeutel}}]{Scheucher2016}%
  \BibitemOpen
  \bibfield  {author} {\bibinfo {author} {\bibfnamefont {M.}~\bibnamefont
  {Scheucher}}, \bibinfo {author} {\bibfnamefont {A.}~\bibnamefont {Hilico}},
  \bibinfo {author} {\bibfnamefont {E.}~\bibnamefont {Will}}, \bibinfo {author}
  {\bibfnamefont {J.}~\bibnamefont {Volz}}, \ and\ \bibinfo {author}
  {\bibfnamefont {A.}~\bibnamefont {Rauschenbeutel}},\ }\href {\doibase
  10.1126/science.aaj2118} {\bibfield  {journal} {\bibinfo  {journal}
  {Science}\ }\textbf {\bibinfo {volume} {354}},\ \bibinfo {pages} {1577}
  (\bibinfo {year} {2016})}\BibitemShut {NoStop}%
\bibitem [{\citenamefont {O'Connor}\ \emph {et~al.}(2014)\citenamefont
  {O'Connor}, \citenamefont {Ginzburg}, \citenamefont
  {Rodr{\'{i}}guez-Fortu{\~{n}}o}, \citenamefont {Wurtz},\ and\ \citenamefont
  {Zayats}}]{OConnor2014}%
  \BibitemOpen
  \bibfield  {author} {\bibinfo {author} {\bibfnamefont {D.}~\bibnamefont
  {O'Connor}}, \bibinfo {author} {\bibfnamefont {P.}~\bibnamefont {Ginzburg}},
  \bibinfo {author} {\bibfnamefont {F.~J.}\ \bibnamefont
  {Rodr{\'{i}}guez-Fortu{\~{n}}o}}, \bibinfo {author} {\bibfnamefont {G.~A.}\
  \bibnamefont {Wurtz}}, \ and\ \bibinfo {author} {\bibfnamefont {A.~V.}\
  \bibnamefont {Zayats}},\ }\href {\doibase 10.1038/ncomms6327} {\bibfield
  {journal} {\bibinfo  {journal} {Nat. Commun.}\ }\textbf {\bibinfo {volume}
  {5}},\ \bibinfo {pages} {5327} (\bibinfo {year} {2014})}\BibitemShut
  {NoStop}%
\bibitem [{\citenamefont {Bliokh}\ \emph {et~al.}(2014)\citenamefont {Bliokh},
  \citenamefont {Bekshaev},\ and\ \citenamefont {Nori}}]{Bliokh2014}%
  \BibitemOpen
  \bibfield  {author} {\bibinfo {author} {\bibfnamefont {K.~Y.}\ \bibnamefont
  {Bliokh}}, \bibinfo {author} {\bibfnamefont {A.~Y.}\ \bibnamefont
  {Bekshaev}}, \ and\ \bibinfo {author} {\bibfnamefont {F.}~\bibnamefont
  {Nori}},\ }\href {\doibase 10.1038/ncomms4300} {\bibfield  {journal}
  {\bibinfo  {journal} {Nat. Commun.}\ }\textbf {\bibinfo {volume} {5}},\
  \bibinfo {pages} {3300} (\bibinfo {year} {2014})}\BibitemShut {NoStop}%
\bibitem [{\citenamefont {Berry}(2009)}]{Berry2009}%
  \BibitemOpen
  \bibfield  {author} {\bibinfo {author} {\bibfnamefont {M.~V.}\ \bibnamefont
  {Berry}},\ }\href {\doibase 10.1088/1464-4258/11/9/094001} {\bibfield
  {journal} {\bibinfo  {journal} {J. Opt. A}\ }\textbf {\bibinfo {volume}
  {11}},\ \bibinfo {pages} {094001} (\bibinfo {year} {2009})}\BibitemShut
  {NoStop}%
\bibitem [{\citenamefont {Bekshaev}\ and\ \citenamefont
  {Soskin}(2007)}]{Bekshaev2007}%
  \BibitemOpen
  \bibfield  {author} {\bibinfo {author} {\bibfnamefont {A.~Y.}\ \bibnamefont
  {Bekshaev}}\ and\ \bibinfo {author} {\bibfnamefont {M.~S.}\ \bibnamefont
  {Soskin}},\ }\href {\doibase 10.1016/j.optcom.2006.10.057} {\bibfield
  {journal} {\bibinfo  {journal} {Opt. Commun.}\ }\textbf {\bibinfo {volume}
  {271}},\ \bibinfo {pages} {332} (\bibinfo {year} {2007})}\BibitemShut
  {NoStop}%
\bibitem [{\citenamefont {Mandel}\ and\ \citenamefont
  {Wolf}(1995)}]{Mandel1995}%
  \BibitemOpen
  \bibfield  {author} {\bibinfo {author} {\bibfnamefont {L.}~\bibnamefont
  {Mandel}}\ and\ \bibinfo {author} {\bibfnamefont {E.}~\bibnamefont {Wolf}},\
  }\href@noop {} {\emph {\bibinfo {title} {{Optical Coherence and Quantum
  Optics}}}},\ \bibinfo {edition} {1st}\ ed.\ (\bibinfo  {publisher} {Cambridge
  University Press},\ \bibinfo {address} {Cambridge},\ \bibinfo {year}
  {1995})\BibitemShut {NoStop}%
\bibitem [{\citenamefont {Jackson}(1999)}]{Jackson1999}%
  \BibitemOpen
  \bibfield  {author} {\bibinfo {author} {\bibfnamefont {J.~D.}\ \bibnamefont
  {Jackson}},\ }\href {\doibase 10.4006/1.3025509} {\emph {\bibinfo {title}
  {{Classical Electrodynamics}}}},\ \bibinfo {edition} {3rd}\ ed.\ (\bibinfo
  {publisher} {Wiley},\ \bibinfo {address} {New York},\ \bibinfo {year}
  {1999})\BibitemShut {NoStop}%
\bibitem [{\citenamefont {Novotny}\ and\ \citenamefont
  {Hecht}(2006)}]{Novotny2006}%
  \BibitemOpen
  \bibfield  {author} {\bibinfo {author} {\bibfnamefont {L.}~\bibnamefont
  {Novotny}}\ and\ \bibinfo {author} {\bibfnamefont {B.}~\bibnamefont
  {Hecht}},\ }\href@noop {} {\emph {\bibinfo {title} {{Principles of
  Nano-Optics}}}},\ \bibinfo {edition} {2nd}\ ed.\ (\bibinfo  {publisher}
  {Cambridge University Press},\ \bibinfo {address} {Cambridge},\ \bibinfo
  {year} {2006})\BibitemShut {NoStop}%
\bibitem [{\citenamefont {Evlyukhin}\ \emph {et~al.}(2012)\citenamefont
  {Evlyukhin}, \citenamefont {Novikov}, \citenamefont {Zywietz}, \citenamefont
  {Eriksen}, \citenamefont {Reinhardt}, \citenamefont {Bozhevolnyi},\ and\
  \citenamefont {Chichkov}}]{Evlyukhin2012}%
  \BibitemOpen
  \bibfield  {author} {\bibinfo {author} {\bibfnamefont {A.~B.}\ \bibnamefont
  {Evlyukhin}}, \bibinfo {author} {\bibfnamefont {S.~M.}\ \bibnamefont
  {Novikov}}, \bibinfo {author} {\bibfnamefont {U.}~\bibnamefont {Zywietz}},
  \bibinfo {author} {\bibfnamefont {R.~L.}\ \bibnamefont {Eriksen}}, \bibinfo
  {author} {\bibfnamefont {C.}~\bibnamefont {Reinhardt}}, \bibinfo {author}
  {\bibfnamefont {S.~I.}\ \bibnamefont {Bozhevolnyi}}, \ and\ \bibinfo {author}
  {\bibfnamefont {B.~N.}\ \bibnamefont {Chichkov}},\ }\href {\doibase
  10.1021/nl301594s} {\bibfield  {journal} {\bibinfo  {journal} {Nano Lett.}\
  }\textbf {\bibinfo {volume} {12}},\ \bibinfo {pages} {3749} (\bibinfo {year}
  {2012})}\BibitemShut {NoStop}%
\bibitem [{\citenamefont {Kuznetsov}\ \emph {et~al.}(2012)\citenamefont
  {Kuznetsov}, \citenamefont {Miroshnichenko}, \citenamefont {Fu},
  \citenamefont {Zhang},\ and\ \citenamefont {Luk'yanchuk}}]{Kuznetsov2012}%
  \BibitemOpen
  \bibfield  {author} {\bibinfo {author} {\bibfnamefont {A.~I.}\ \bibnamefont
  {Kuznetsov}}, \bibinfo {author} {\bibfnamefont {A.~E.}\ \bibnamefont
  {Miroshnichenko}}, \bibinfo {author} {\bibfnamefont {Y.~H.}\ \bibnamefont
  {Fu}}, \bibinfo {author} {\bibfnamefont {J.}~\bibnamefont {Zhang}}, \ and\
  \bibinfo {author} {\bibfnamefont {B.}~\bibnamefont {Luk'yanchuk}},\ }\href
  {\doibase 10.1038/srep00492} {\bibfield  {journal} {\bibinfo  {journal} {Sci.
  Rep.}\ }\textbf {\bibinfo {volume} {2}},\ \bibinfo {pages} {492} (\bibinfo
  {year} {2012})}\BibitemShut {NoStop}%
\bibitem [{\citenamefont {Krasnok}\ \emph {et~al.}(2012)\citenamefont
  {Krasnok}, \citenamefont {Miroshnichenko}, \citenamefont {Belov},\ and\
  \citenamefont {Kivshar}}]{Krasnok2012}%
  \BibitemOpen
  \bibfield  {author} {\bibinfo {author} {\bibfnamefont {A.~E.}\ \bibnamefont
  {Krasnok}}, \bibinfo {author} {\bibfnamefont {A.~E.}\ \bibnamefont
  {Miroshnichenko}}, \bibinfo {author} {\bibfnamefont {P.~A.}\ \bibnamefont
  {Belov}}, \ and\ \bibinfo {author} {\bibfnamefont {Y.~S.}\ \bibnamefont
  {Kivshar}},\ }\href {\doibase 10.1063/1.4750083} {\bibfield  {journal}
  {\bibinfo  {journal} {Opt. Express}\ }\textbf {\bibinfo {volume} {20}},\
  \bibinfo {pages} {20599} (\bibinfo {year} {2012})}\BibitemShut {NoStop}%
\bibitem [{\citenamefont {Mie}(1908)}]{Mie1908}%
  \BibitemOpen
  \bibfield  {author} {\bibinfo {author} {\bibfnamefont {G.}~\bibnamefont
  {Mie}},\ }\href@noop {} {\bibfield  {journal} {\bibinfo  {journal} {Ann.
  Phys. (Berlin)}\ }\textbf {\bibinfo {volume} {4}},\ \bibinfo {pages} {377}
  (\bibinfo {year} {1908})}\BibitemShut {NoStop}%
\bibitem [{\citenamefont {Wo{\'{z}}niak}\ \emph {et~al.}(2015)\citenamefont
  {Wo{\'{z}}niak}, \citenamefont {Banzer},\ and\ \citenamefont
  {Leuchs}}]{Wozniak2015}%
  \BibitemOpen
  \bibfield  {author} {\bibinfo {author} {\bibfnamefont {P.}~\bibnamefont
  {Wo{\'{z}}niak}}, \bibinfo {author} {\bibfnamefont {P.}~\bibnamefont
  {Banzer}}, \ and\ \bibinfo {author} {\bibfnamefont {G.}~\bibnamefont
  {Leuchs}},\ }\href {\doibase 10.1002/lpor.201400188} {\bibfield  {journal}
  {\bibinfo  {journal} {Laser Photon. Rev.}\ }\textbf {\bibinfo {volume} {9}},\
  \bibinfo {pages} {231} (\bibinfo {year} {2015})}\BibitemShut {NoStop}%
\bibitem [{\citenamefont {Neugebauer}\ \emph {et~al.}(2016)\citenamefont
  {Neugebauer}, \citenamefont {Wo{\'{z}}niak}, \citenamefont {Bag},
  \citenamefont {Leuchs},\ and\ \citenamefont {Banzer}}]{Neugebauer2016}%
  \BibitemOpen
  \bibfield  {author} {\bibinfo {author} {\bibfnamefont {M.}~\bibnamefont
  {Neugebauer}}, \bibinfo {author} {\bibfnamefont {P.}~\bibnamefont
  {Wo{\'{z}}niak}}, \bibinfo {author} {\bibfnamefont {A.}~\bibnamefont {Bag}},
  \bibinfo {author} {\bibfnamefont {G.}~\bibnamefont {Leuchs}}, \ and\ \bibinfo
  {author} {\bibfnamefont {P.}~\bibnamefont {Banzer}},\ }\href {\doibase
  10.1038/ncomms11286 OPEN} {\bibfield  {journal} {\bibinfo  {journal} {Nat.
  Commun.}\ }\textbf {\bibinfo {volume} {7}},\ \bibinfo {pages} {11286}
  (\bibinfo {year} {2016})}\BibitemShut {NoStop}%
\bibitem [{\citenamefont {Lukosz}\ and\ \citenamefont
  {Kunz}(1977)}]{Lukosz1977b}%
  \BibitemOpen
  \bibfield  {author} {\bibinfo {author} {\bibfnamefont {W.}~\bibnamefont
  {Lukosz}}\ and\ \bibinfo {author} {\bibfnamefont {R.~E.}\ \bibnamefont
  {Kunz}},\ }\href@noop {} {\bibfield  {journal} {\bibinfo  {journal} {J. Opt.
  Soc. Am.}\ }\textbf {\bibinfo {volume} {67}},\ \bibinfo {pages} {1615}
  (\bibinfo {year} {1977})}\BibitemShut {NoStop}%
\bibitem [{\citenamefont {Courtois}\ \emph {et~al.}(1996)\citenamefont
  {Courtois}, \citenamefont {Courty},\ and\ \citenamefont
  {Mertz}}]{Courtois1996}%
  \BibitemOpen
  \bibfield  {author} {\bibinfo {author} {\bibfnamefont {J.}~\bibnamefont
  {Courtois}}, \bibinfo {author} {\bibfnamefont {J.}~\bibnamefont {Courty}}, \
  and\ \bibinfo {author} {\bibfnamefont {J.}~\bibnamefont {Mertz}},\ }\href
  {\doibase 10.1103/PhysRevA.53.1862} {\bibfield  {journal} {\bibinfo
  {journal} {Phys. Rev. A}\ }\textbf {\bibinfo {volume} {53}},\ \bibinfo
  {pages} {1862} (\bibinfo {year} {1996})}\BibitemShut {NoStop}%
\bibitem [{\citenamefont {{Le Kien}}\ and\ \citenamefont
  {Rauschenbeutel}(2016)}]{LeKien2016}%
  \BibitemOpen
  \bibfield  {author} {\bibinfo {author} {\bibfnamefont {F.}~\bibnamefont {{Le
  Kien}}}\ and\ \bibinfo {author} {\bibfnamefont {A.}~\bibnamefont
  {Rauschenbeutel}},\ }\href {\doibase 10.1103/PhysRevA.93.043828} {\bibfield
  {journal} {\bibinfo  {journal} {Phys. Rev. A}\ }\textbf {\bibinfo {volume}
  {93}},\ \bibinfo {pages} {1} (\bibinfo {year} {2016})}\BibitemShut {NoStop}%
\bibitem [{\citenamefont {Richards}\ and\ \citenamefont
  {Wolf}(1959)}]{Richards1959}%
  \BibitemOpen
  \bibfield  {author} {\bibinfo {author} {\bibfnamefont {B.}~\bibnamefont
  {Richards}}\ and\ \bibinfo {author} {\bibfnamefont {E.}~\bibnamefont
  {Wolf}},\ }\href {\doibase 10.1098/rspa.1959.0200} {\bibfield  {journal}
  {\bibinfo  {journal} {Proc. R. Soc. A}\ }\textbf {\bibinfo {volume} {253}},\
  \bibinfo {pages} {358} (\bibinfo {year} {1959})}\BibitemShut {NoStop}%
\bibitem [{\citenamefont {Youngworth}\ and\ \citenamefont
  {Brown}(2000)}]{Youngworth2000}%
  \BibitemOpen
  \bibfield  {author} {\bibinfo {author} {\bibfnamefont {K.}~\bibnamefont
  {Youngworth}}\ and\ \bibinfo {author} {\bibfnamefont {T.}~\bibnamefont
  {Brown}},\ }\href {\doibase 10.1364/OE.7.000077} {\bibfield  {journal}
  {\bibinfo  {journal} {Opt. Express}\ }\textbf {\bibinfo {volume} {7}},\
  \bibinfo {pages} {77} (\bibinfo {year} {2000})}\BibitemShut {NoStop}%
\bibitem [{\citenamefont {Antognozzi}\ \emph {et~al.}(2016)\citenamefont
  {Antognozzi}, \citenamefont {Bermingham}, \citenamefont {Harniman},
  \citenamefont {Simpson}, \citenamefont {Senior}, \citenamefont {Hayward},
  \citenamefont {Hoerber}, \citenamefont {Dennis}, \citenamefont {Bekshaev},
  \citenamefont {Bliokh},\ and\ \citenamefont {Nori}}]{Antognozzi2016}%
  \BibitemOpen
  \bibfield  {author} {\bibinfo {author} {\bibfnamefont {M.}~\bibnamefont
  {Antognozzi}}, \bibinfo {author} {\bibfnamefont {C.~R.}\ \bibnamefont
  {Bermingham}}, \bibinfo {author} {\bibfnamefont {R.}~\bibnamefont
  {Harniman}}, \bibinfo {author} {\bibfnamefont {S.}~\bibnamefont {Simpson}},
  \bibinfo {author} {\bibfnamefont {J.}~\bibnamefont {Senior}}, \bibinfo
  {author} {\bibfnamefont {R.}~\bibnamefont {Hayward}}, \bibinfo {author}
  {\bibfnamefont {H.}~\bibnamefont {Hoerber}}, \bibinfo {author} {\bibfnamefont
  {M.~R.}\ \bibnamefont {Dennis}}, \bibinfo {author} {\bibfnamefont {A.~Y.}\
  \bibnamefont {Bekshaev}}, \bibinfo {author} {\bibfnamefont {K.~Y.}\
  \bibnamefont {Bliokh}}, \ and\ \bibinfo {author} {\bibfnamefont
  {F.}~\bibnamefont {Nori}},\ }\href {\doibase 10.13140/RG.2.1.3724.8166}
  {\bibfield  {journal} {\bibinfo  {journal} {Nat. Phys.}\ }\textbf {\bibinfo
  {volume} {12}},\ \bibinfo {pages} {731} (\bibinfo {year} {2016})}\BibitemShut
  {NoStop}%
\bibitem [{\citenamefont {Picardi}\ \emph {et~al.}(2017)\citenamefont
  {Picardi}, \citenamefont {Manjavacas}, \citenamefont {Zayats},\ and\
  \citenamefont
  {Rodr{\'{i}}guez-Fortu{\~{n}}o}}]{MichelaF.PicardiAlejandroManjavacasAnatolyV.Zayats2017}%
  \BibitemOpen
  \bibfield  {author} {\bibinfo {author} {\bibfnamefont {M.~F.}\ \bibnamefont
  {Picardi}}, \bibinfo {author} {\bibfnamefont {A.}~\bibnamefont {Manjavacas}},
  \bibinfo {author} {\bibfnamefont {A.~V.}\ \bibnamefont {Zayats}}, \ and\
  \bibinfo {author} {\bibfnamefont {F.~J.}\ \bibnamefont
  {Rodr{\'{i}}guez-Fortu{\~{n}}o}},\ }\href {\doibase
  10.1103/PhysRevB.95.245416} {\bibfield  {journal} {\bibinfo  {journal} {Phys.
  Rev. B}\ }\textbf {\bibinfo {volume} {95}},\ \bibinfo {pages} {245416}
  (\bibinfo {year} {2017})}\BibitemShut {NoStop}%
\bibitem [{\citenamefont {Wang}\ \emph {et~al.}(2017)\citenamefont {Wang},
  \citenamefont {Zhang}, \citenamefont {Kovalevitch}, \citenamefont {Salut},
  \citenamefont {Kim}, \citenamefont {Suarez}, \citenamefont {Bernal},
  \citenamefont {Herzig}, \citenamefont {Lu},\ and\ \citenamefont
  {Grosjean}}]{Wang2017}%
  \BibitemOpen
  \bibfield  {author} {\bibinfo {author} {\bibfnamefont {M.}~\bibnamefont
  {Wang}}, \bibinfo {author} {\bibfnamefont {H.}~\bibnamefont {Zhang}},
  \bibinfo {author} {\bibfnamefont {T.}~\bibnamefont {Kovalevitch}}, \bibinfo
  {author} {\bibfnamefont {R.}~\bibnamefont {Salut}}, \bibinfo {author}
  {\bibfnamefont {M.-S.}\ \bibnamefont {Kim}}, \bibinfo {author} {\bibfnamefont
  {A.}~\bibnamefont {Suarez}}, \bibinfo {author} {\bibfnamefont {M.-P.}\
  \bibnamefont {Bernal}}, \bibinfo {author} {\bibfnamefont {H.-P.}\
  \bibnamefont {Herzig}}, \bibinfo {author} {\bibfnamefont {H.}~\bibnamefont
  {Lu}}, \ and\ \bibinfo {author} {\bibfnamefont {T.}~\bibnamefont
  {Grosjean}},\ }\href {https://arxiv.org/abs/1710.03584} {\bibfield  {journal}
  {\bibinfo  {journal} {arXiv: 1710.03584}\ } (\bibinfo {year}
  {2017})}\BibitemShut {NoStop}%
\end{thebibliography}%
\end{document}